\documentclass[%
 prl,preprint,groupedaddress,
 amsmath,amssymb,
 aps,
]{revtex4-1}

\usepackage{graphicx}
\usepackage{dcolumn}
\usepackage{bm}

\begin{document}


\title{Flavor-Tuned 125~GeV SUSY Higgs Boson at the LHC :\\
MSSM and NATURAL SUSY TESTS}

\author{Vernon Barger$^*$, Peisi Huang$^*$, Muneyuki Ishida$^\dagger$, and Wai-Yee Keung$^\ddagger$}
\affiliation{
$^*$ Department of Physics, University of Wisconsin, Madison, WI 53706, USA\\
$^\dagger$ Department of Physics, Meisei University, Hino, Tokyo 191-8506, Japan\\
$^\ddagger$ Department of Physics, University of Illinois at Chicago, IL 60607, USA
}%





\date{\today}

\begin{abstract}
We show that an enhanced two-photon signal of the Higgs boson, $h$,
observed with 125~GeV mass by the ATLAS and CMS collaborations,
can be obtained if it is identified principally with the neutral $H_u^0$ of the two Higgs doublets of minimal Supersymmetry.
We focus on sparticles and the pseudoscalar Higgs $A$ at the TeV scale.
The off-diagonal element of the ($H_u^0$,$H_d^0$) mass matrix in the flavor basis must be suppressed, and this requires
both a large Higgsino mass parameter, $\mu\sim$TeV, and large tan$\beta$.
A MSSM sum rule is derived that relates $\gamma\gamma$ and $b\bar b$ rates, and a
$\gamma\gamma$ enhancement relative to the SM predicts the $b\bar b$ reduction.
On the contrary,
Natural SUSY requires $|\mu|\stackrel{<}{\scriptscriptstyle \sim}0.5$~TeV,
for which $\gamma\gamma$ is reduced and $b\bar b$ is enhanced.
This conclusion is independent of the $m_A$-value and the SUSY quantum
correction $\Delta_b$.
Relative $\tau\bar\tau$ to $b\bar b$ rates are sensitive to $\Delta_b$.
\end{abstract}

\pacs{14.80.Ly 12.60.Jv}
\maketitle

A $\gamma\gamma$ enhancement of the 125 GeV Higgs boson signal relative to
the Standard Model (SM) expectation has been reported
by the ATLAS and CMS experiments at the LHC\cite{ATLAS,CMS}.
We investigate this in the Minimal Supersymmetric Standard Model (MSSM)
in the region of large
$m_A\sim$TeV by flavor-tuning of the mixing angle $\alpha$ between two neutral
CP-even Higgs flavor states $H_u^0$ and $H_d^0$, with the 125 GeV Higgs signal identified
principally with $H_u^0$.
Then, the $b\bar b$ decay, which is predicted to be dominant decay
of the SM Higgs boson, is reduced and
the cross sections of the other channels are correspondingly enhanced,
except possibly the $\tau\tau$ channel.
We relate the cross-section enhancements$/$suppressions in
$\gamma\gamma /b\bar b/\tau\tau$
channels compared with those of the SM Higgs boson.
We also consider the consequences for Natural SUSY\cite{NS}.
Our focus is on a heavy pseudoscalar $A$ and large
$\tan\beta \equiv \langle H_u^0\rangle/\langle H_d^0\rangle$, 
a region that has not yet been constrained by LHC experiments\cite{tanb}.
Light stau\cite{stau,hagiwara} and light stop\cite{falk} scenarios that have been considered
are outside of our purview.

\noindent\underline{\it Ratios of the SUSY Higgs couplings to those of the SM Higgs}\ \ \
The SUSY Higgs mechanism is based on the two Higgs doublet model
of type II\cite{BHG,Chung,Hunter}
 with the $H_u$ doublet coupled to up-type quarks and the $H_d$ doublet coupled
to down-type quarks.
After spontaneous symmetry breaking, the physical Higgs states are two CP-even neutral
Higgs $h,H$, one CP-odd neutral pseudo-scalar $A$ and the charged Higgs $H^\pm$.

We focus on the CP-even neutral Higgs boson $h$ and $H$, which are related to the
flavor eigenstates $H_u^0$ and $H_d^0$ by
\begin{eqnarray}
\frac{h}{\sqrt 2} &=& c_\alpha H_u^0 -s_\alpha H_d^0\ ,\ \ \
\frac{H}{\sqrt 2} = s_\alpha H_u^0 +c_\alpha H_d^0\ ,
\label{eq1}
\end{eqnarray}
where $H_{u,d}^0$  is the shorthand for
the real part of $H_{u,d}^0 - \langle H_{u,d}^0\rangle $.
We use the notation  $s_\alpha ={\rm sin}\alpha$, $c_\alpha ={\rm cos}\alpha$,
and $t_\alpha ={\rm tan}\alpha$.
Our interest is in large tan$\beta$,
tan$\beta \stackrel{>}{\scriptscriptstyle \sim} 20$, and in the
decoupling regime with large $m_A$ for which
$\alpha\simeq \beta-\frac{\pi}{2}$.

The ratios of the $h$ and $H$ couplings to those of the SM Higgs $h_{SM}$,
denoted as $r_{PP}^{h,H}(\equiv g_{h,H\ P\bar P}/g_{h_{SM} P\bar P})$, are given by
\begin{eqnarray}
r_{VV}^h &=& s_{\beta -\alpha},\ \ \ \  r_{tt}^h = r_{cc}^h = \frac{c_\alpha}{s_\beta},\ \  r_{\tau\tau}^h = \frac{-s_\alpha}{c_\beta},\ \
r_{bb}^h = \frac{-s_\alpha}{c_\beta}\left[ 1-\frac{\Delta_{b}}{1+\Delta_{b}}
(1+\frac{1}{t_\alpha t_\beta})   \right]\ \ \nonumber\\
r_{VV}^H &=& c_{\beta -\alpha},\ \ \ \  r_{tt}^H = r_{cc}^H = \frac{s_\alpha}{s_\beta},\ \ r_{\tau\tau}^H = \frac{c_\alpha}{c_\beta},\ \
r_{bb}^H = \frac{c_\alpha}{c_\beta}\left[ 1-\frac{\Delta_{b}}{1+\Delta_{b}}
(1- \frac{t_\alpha}{t_\beta} )   \right]
\label{eq2}
\end{eqnarray}
where we include the 1-loop contribution $\Delta_b$ to the $b\bar b$ coupling.
It is the
$b$-quark mass correction factor \cite{Dothers,Djouadi}, which may be sizable,
especially if both $\mu$ and tan$\beta$ are large.
\begin{eqnarray}
\Delta_b &=& \bar\mu\ t_\beta\left[
\frac{2\alpha_s}{3\pi} \hat m_{\tilde g}
 I(\hat m_{\tilde g}^2,\hat m_{\tilde b_1}^2,\hat m_{\tilde b_2}^2)
 + \frac{h_t^2}{16\pi^2}a_t I(\bar\mu^2, \hat m_{\tilde t_1}^2,\hat m_{\tilde t_2}^2) \right]
\label{eq3}\\
 && I(x,y,z) = -\frac{xy\ {\rm ln}x/y+yz\ {\rm ln}y/z+zx\ {\rm ln}z/x}{(x-y)(y-z)(z-x)}
\nonumber\\
 &&  I(x,y,z=y) = -\left[ x-y + x {\rm log}\frac{y}{x} \right] / (x-y)^2\ ,\ \ \ \
      I(x,x,x) = \frac{1}{2x} \ .
\label{eq4}
\end{eqnarray}
The first(second) term of $\Delta_b$ is due to the sbottom-gluino(stop-chargino) loop.
We take $M_{\rm susy}=1$~TeV and sparticle masses $\hat m$
in units of $M_{\rm susy}$.
The top Yukawa coupling is $h_t=\bar m_t/v_u=\bar m_t/(v\ s_\beta)$ and
$\bar m_t=m_t(\bar m_t)=163.5$~GeV is the running top quark mass\cite{Langacker}.
We consider $m_Q=m_U=m_D=M_{\rm susy}$ for the squark masses in the third generation.

The off-diagonal element of the stop squared mass matrix is $\bar m_t X_t$ where
the stop mixing parameter $X_t$ is given by $X_t = A_t - \mu/t_\beta$.
These quantities are also defined in units of $M_{\rm susy}$ as
$a_t\equiv A_t/M_{\rm susy}$, $\bar\mu\equiv\mu/M_{\rm susy}$, and
$x_t\equiv X_t/M_{\rm susy}=a_t - \bar\mu/t_\beta$. Our sign convention for $\mu$ and $A_t$ is the same as \cite{CarenaM}, which are opposite sign convention of \cite{wss}.
We fix $\hat m_{\tilde g}=2$, well above the current LHC reach,
$\hat m_{\tilde b_1}=\hat m_{\tilde b_2}=1$, and $\hat m_{\tilde t_1}=0.8$,
 $\hat m_{\tilde t_2}=1.2$. A stop mass difference
$m_{\tilde t_2}-m_{\tilde t_1}\ge 0.4$~TeV is chosen in accord with
Natural SUSY prediction\cite{Hmass}.
Then $\Delta_b$ is well approximated numerically by
\begin{eqnarray}
\Delta_b &\simeq & \bar\mu\frac{t_\beta}{20}\left[0.26 + \left(\frac{0.09}{|\bar\mu|+0.6}-0.003\right) a_t\right]\ ,
\label{eq5}
\end{eqnarray}
where the first and the second terms in the square bracket are the numerical values from the
gluino and the chargino contributions respectively. 

The chargino and neutralino masses have no special role except possibly in
$b\rightarrow s\gamma$ decay, but consistency with Natural SUSY
has been found there\cite{BBM}. Large $m_A$ implies large charged Higgs $H^+$ mass
that suppresses the $H^+$ loop contribution to $b\rightarrow s\gamma$.

The $gg,\gamma\gamma$ coupling ratios $r_{gg,\gamma\gamma}^{\phi}$ for $\phi =h,H,A$
 relative to those of $h_{SM}$ are \cite{total}
\begin{eqnarray}
r_{gg}^\phi &=&
\frac{I_{tt}^\phi r_{tt}^h + I_{bb}^\phi r_{bb}^h}{I_{tt}^\phi +I_{bb}^\phi}\ ,
\ \ \ \ \ \
r_{\gamma\gamma}^\phi =
  \frac{\frac{7}{4}I_{WW}^\phi\ r_{VV}^h-\frac{4}{9}I_{tt}^\phi r_{tt}^h-\frac{1}{9}I_{bb}^\phi r_{bb}^h}{\frac{7}{4}I_{WW}^\phi-\frac{4}{9}I_{tt}^\phi -\frac{1}{9}I_{bb}^\phi} \ ,
\label{eq8}
\end{eqnarray}
where $I^\phi_{WW,tt,bb}$ represent the triangle-loop contributions to the amplitudes
normalized to the $m_h\rightarrow 0$ limit\cite{book,dilaton,Kingman1}.

The $XX\rightarrow h \rightarrow PP$ cross section ratios\cite{total}
 relative to $h_{SM}$ are obtained from
\begin{eqnarray}
\sigma_P &\equiv & \frac{\sigma_{PP}}{\sigma_{\rm SM}} = \frac{\sigma_{XX\rightarrow PP}}{\sigma_{XX\rightarrow h_{SM}\rightarrow PP}}
 = \frac{|r_{XX}^h r_{PP}^h|^2}{R^h} \ ,\ \ \
\label{eq9}\\
R^{h} &=& \frac{\Gamma_{\rm tot}^h}{\Gamma_{\rm tot}^{h_{SM}}} = 0.57|r_{bb}^{h}|^2+0.06|r_{\tau\tau}^{h}|^2+0.25|r_{VV}^{h}|^2
+0.09|r_{gg}^{h}|^2+0.03|r_{cc}^{h}|^2\ ,
\label{eq10}
\end{eqnarray}
where $R^h$ is the ratio of the $h$ total width to that of $h_{SM}$,
$\Gamma_{h_{SM}}^{\rm tot}=4.14$~MeV\cite{Heine} for $m_h=125.5$~GeV.
The coefficients in Eq.~(\ref{eq10}) are the SM Higgs branching fractions.
Here we have assumed no appreciable decays to dark matter.

\noindent\underline{\it Sum rule of cross-section ratios}\ \ \
In the large $m_A$ region close to the decoupling limit, $\alpha$ takes a value
\begin{eqnarray}
\alpha &=& \beta-\frac{\pi}{2}+\epsilon
\label{eq11}
\end{eqnarray}
with $|\epsilon|<\frac{\pi}{2}-\beta$.
Then, the $r_{XX}^h$ of Eq.~(\ref{eq2}) are well approximated by
\begin{eqnarray}
r_{VV}^{h} &=& 1,\ \ r_{tt,cc}^{h}=1+\epsilon/t_\beta ,\ \
r_{\tau\tau}^h \simeq 1-\epsilon t_\beta ,\ \
r_{bb}^h \simeq 1-\frac{1}{1+\Delta_{b}}\epsilon t_\beta \ .
\label{eq12}
\end{eqnarray}
through first order in $\epsilon$.
The $r_{tt,cc}^{h}$ are close to unity because those deviations from SM are
$t_\beta$ suppressed. Thus,
\begin{eqnarray}
r_{gg}^h &\simeq& r_{\gamma\gamma}^h \simeq 1\ ,
\label{eq13}
\end{eqnarray}
since the bottom triangle loop function $I_{bb}^h$ is negligible in Eq.~(\ref{eq8}).
Only $r_{bb}^h,r_{\tau\tau}^h$
can deviate sizably from unity for large $m_A$ and large tan$\beta$.
Following Eqs.~(\ref{eq9}) and (\ref{eq10}), the
$\sigma_P\equiv\sigma_{PP}/\sigma_{\rm SM}$
of the other channels
are commonly reduced(enhanced) in correspondence with $r_{bb}^h>1$ ($r_{bb}^h<1$).
We predict the cross sections relative to their individual SM expectations
\begin{eqnarray}
\sigma_\gamma = \sigma_W =\sigma_Z
 && = \frac{1}{0.6 (r_{bb}^{h})^2 + 0.4}\ ,\ \ \
\label{eq14}
\end{eqnarray}
and
\begin{eqnarray}
 0.4 \sigma_{\gamma} + 0.6 \sigma_{b}  &=& 1
\label{eq14b}
\end{eqnarray}
where the SM $b\bar b$ branching fraction \cite{hdecay} is approximated as 60\% .
Equation (\ref{eq14}) holds independently of the production process.
Enhanced $\sigma_\gamma$ implies reduced $\sigma_{b}$,
as well as enhanced $\sigma_W$ and $\sigma_Z$. 

\noindent\underline{\it Flavor-Tuning of mixing angle $\alpha$}\ \ \
Note that $r_{bb,\tau\tau}^h=1$ in the exact decoupling limit $m_A\rightarrow\infty$
for which $\epsilon=0$.
Flavor-tuning of $\epsilon$ to be small but non-zero is necessary to obtain a significant variation of $r_{bb}^h$
from unity.
Positive(negative) $\epsilon$ gives $bb$-reduction(enhancement).

The mixing angle $\alpha$ is obtained by diagonalizing the squared-mass matrix of the neutral Higgs
in the $u,d$ basis. Their elements at tree-level are
\begin{eqnarray}
 ({M}_{ij}^{2})^{\rm tree} &=& M_Z^2s_\beta^2+m_A^2c_\beta^2;\ \ \
M_Z^2 c_\beta^2+m_A^2 s_\beta^2;\ \ \  -(M_Z^2 +m_A^2) s_\beta c_\beta
\label{eq15}
\end{eqnarray}
for $ij=11;22;12$, respectively, which gives $\epsilon < 0$ in all region of $m_A$.
Thus, in order to get $b\bar b$-reduction, it is necessary to cancel
$(M_{12}^2)^{\rm tree}$ by higher order terms $\Delta M_{ij}^2$.

In the 2-loop leading-log(LL) approximation the $\Delta M_{ij}^2$
are given \cite{CarenaM,HHM} by
\begin{eqnarray}
{M}_{ij}^2 &=& ({M}_{ij}^{2})^{\rm tree} + \Delta{M}_{ij}^2
\label{eq16}
\end{eqnarray}
where
\begin{eqnarray}
\Delta{M}_{11}^2 &=&
F_3 \frac{3 \bar m_t^4}{4\pi^2 v^2 s_\beta^2 }
\left[ t(1- G_\frac{15}{2} t) + a_t x_t (1-\frac{a_t x_t}{12})(1- 2 G_\frac{9}{2} t) \right] -M_Z^2s_\beta^2 (1-F_3)
\nonumber \\
\Delta M_{22}^2 &=& - F_\frac{3}{2} \frac{ \bar m_t^4}{16\pi^2 v^2 s_\beta^2}
\left[(1 - 2G_\frac{9}{2} t) (x_t \bar\mu )^2 \right] \\ \nonumber
\Delta M_{12}^2 &=& - F_\frac{9}{4}\frac{3 \bar m_t^4}{8\pi^2 v^2 s_\beta^2 }
\left[ (1 - 2G_\frac{9}{2} t) (x_t \bar\mu )(1-\frac{a_t x_t}{6}) \right]
 +M_Z^2 s_\beta c_\beta (1-F_\frac{3}{2})
\label{eq17}
\end{eqnarray}
where $F_l = 1/(1+ \frac{h_t^2}{8\pi^2}t)$ with
$l=3,\frac{3}{2},\frac{9}{4}$ and
$G_l =- \frac{1}{16\pi^2}(l h_t^2-32\pi\alpha_s)$ with
$l=\frac{15}{2},\frac{9}{2}$.
The $F_l$  are due to the wave function (WF) renormalization of the $H_u$ field
and the index $l$ is  numbers of $H_u^0$ fields in the
effective potential of the two Higgs doublet model.
$F_3 \xi^{4}\simeq F_\frac{9}{4} \xi^{3}\simeq F_\frac{3}{2} \xi^{2}\simeq 1$ where
$\xi$ is defined by
$H_u(M_s)=H_u(\bar m_t)\xi$ where $\xi=F_\frac{3}{4}^{-1}$.

The parameter tan$\beta = v_u/v_d$ is defined in terms of the Higgs vacuum
expectation values
 $v_{u,d}=\langle H_{u,d}^0\rangle$ at
the minimum of the 1-loop effective potential at the weak scale $\mu=\bar m_t$
and $v=\sqrt{v_u^2+v_d^2}\simeq 174$~GeV,
while $a_t,x_t,\bar\mu$ have scale $\mu = M_{\rm susy}$.
The relation
 cot$\beta (\bar m_t)=$cot$\beta (M_s)\ \xi^{-1}$ will be used
in the following calculation.

Numerically $\alpha_s=\alpha_s(\bar m_t)=0.109$ giving $-32\pi\alpha_s=-10.9$,
while $h_t=\bar m_t/v=0.939$ is small. $G_{\frac{15}{2},\frac{9}{2}}=0.0274,0.0442$ and
$t=$log$(\frac{1~{\rm TeV}}{\bar m_t})^2=3.62$; thus, $G_\frac{15}{2}t=0.099$ and $2G_\frac{9}{2}t=0.320$, and $F_3=0.892$.

\begin{figure}[htb]
\begin{center}
\resizebox{0.7\textwidth}{!}{
  \includegraphics{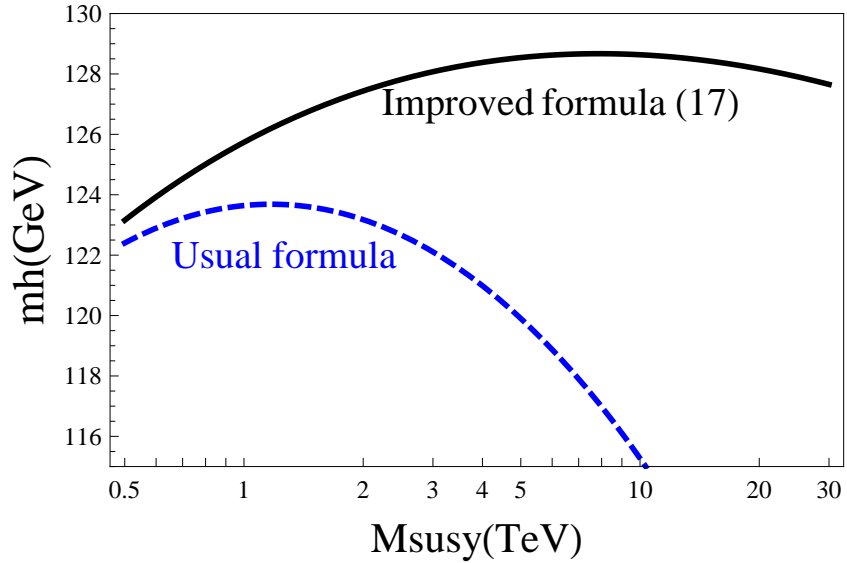}
}
\end{center}
\caption{$M_{\rm susy}$ dependence of Higgs mass $m_h$
by the improved formula Eq.~(\ref{eq17}) (solid black)
in comparison with the one by the usual 2LL approximation (dashed blue) where the first $F_3$ is removed and
$G_\frac{15}{2}$ and $G_\frac{9}{2}$ are replaced by $G_\frac{3}{2}$.
In this illustration $x_t$ is taken to be $\sqrt 6$ following the "maximal-mixing" condition, and tan$\beta=20$. }
\label{fig2}
\end{figure}

In large $m_A$ limit, the $m_h^2$ expression is
\begin{eqnarray}
 m_h^2 &=& M_Z^2c_{2\beta}^2 + F_3 \frac{3\bar m_t^4}{4\pi^2 v^2}\left[
t(1-G_\frac{15}{2} t)+(1-2G_\frac{9}{2}t)(x_t^2-\frac{x_t^4}{12})  \right]
\nonumber \\
&& \qquad\qquad
-M_Z^2 [s_\beta^4 (1-F_3) -2 s_\beta^2 c_\beta^2 (1-F_\frac{3}{2})]
\label{eq18}
\end{eqnarray}
where the Higgs WF renormalization factor $\xi$ is retained in the denominator of $F_3$.
This $F_3$ factor is usually expanded to the numerator in 2LL approximation, and
correspondingly $G_\frac{15}{2}$ and $G_\frac{9}{2}$ are replaced by $G_\frac{3}{2}$:
$m_h^2=M_Z^2c_{2\beta}^2+\frac{3\bar m_t^4}{4\pi^2v^2}[t(1-G_3t)+(1-2G_3t)(x_t^2-\frac{x_t^4}{12})]-M_Z^2s_\beta^4\frac{3h_t^2}{8\pi^2}t$.\ \ \ 
However, numerically Eq.~(\ref{eq18}) significantly increases $m_h$
at large $M_{\rm susy}$ as shown in Fig.~\ref{fig2}:
Eq.~(\ref{eq18}) gives increasing $m_h$ as $M_{\rm susy}$ increases up to $\sim 7$ TeV,
while the usual formula with the expansion approximated for $F_3$
gives decreasing $m_h$ when $M_{\rm susy}>1.3$~TeV
and is not applicable at large $M_{\rm susy}$.

The experimental $m_h$ determinations from the LHC experiments are\cite{ATLAS,CMS}
\begin{eqnarray}
m_h &=& 125.3\pm 0.4\pm 0.4,\ \ \ \ 126.0\pm 0.4\pm 0.4~{\rm GeV}
\label{eq19}
\end{eqnarray}
It seems unlikely that the central $m_h$ determination will change much with
larger statistics because of the excellent mass resolution in the $\gamma\gamma$ channel.
The experimental $m_h$ value is near the maximum possible value of $m_h$
in Eq.~(\ref{eq18}) and this constrains the value of $x_t$
to $|x_t|\simeq \sqrt 6$, to maximize the term $x_t^2-\frac{x_t^4}{12}$.
This is known as "maximal-mixing" in the stop mass-matrix\cite{Hmass}.
In Eq.~(\ref{eq18}) we require $m_h\ge 124$~GeV. This implies
\begin{eqnarray}
 1.95(\equiv x_{\rm tmin})\ \  <\ \  |x_t|\ \  <\ \  2.86(\equiv x_{\rm tmax})\ ,
\label{eq20}
\end{eqnarray}
where we should note that
the positive $x_t$ branch is favored by the SUSY renormalization group prediction\cite{Hmass}.

\begin{figure}[htb]
\begin{center}
\resizebox{0.7\textwidth}{!}{
  \includegraphics{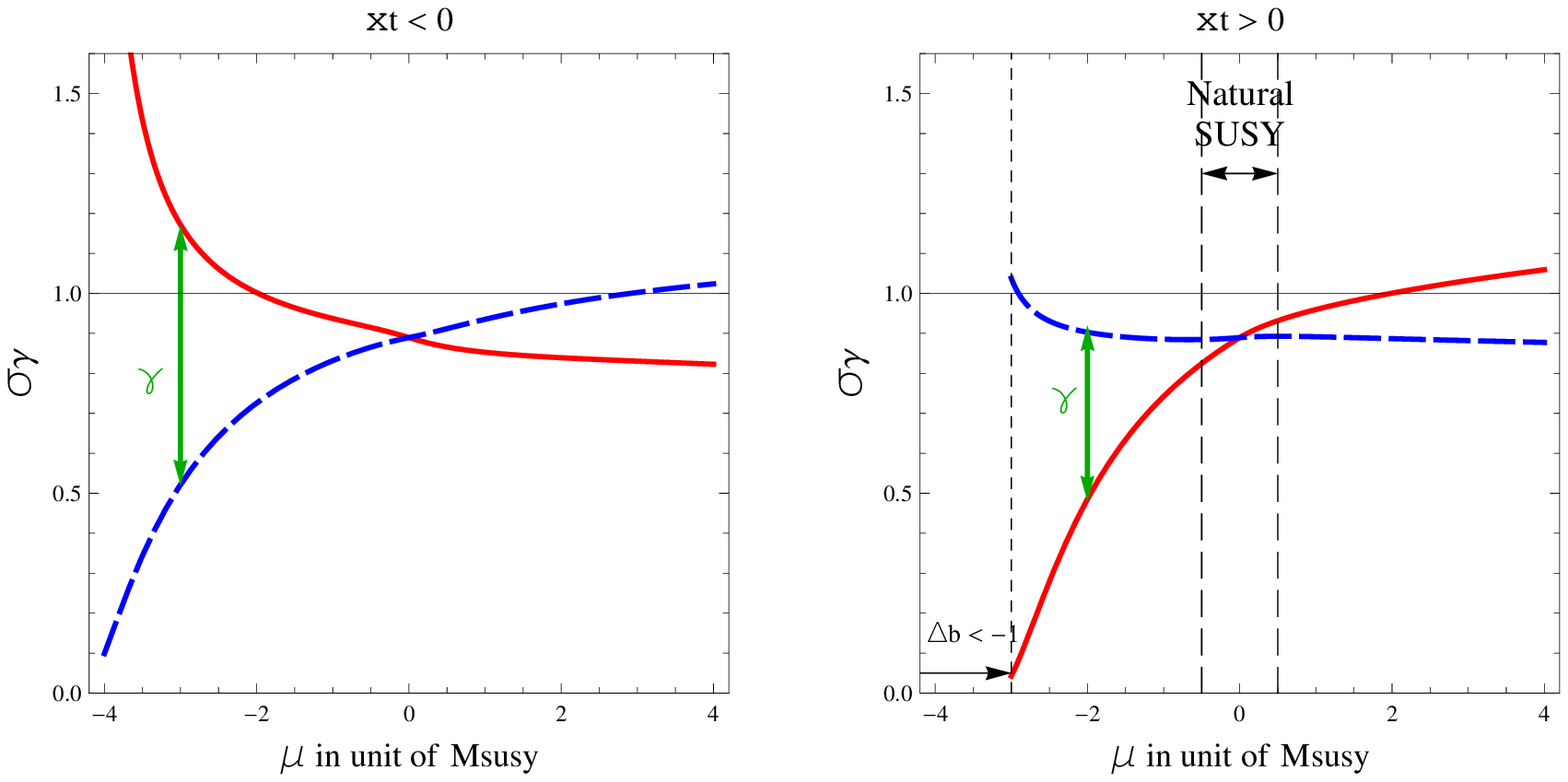}
}\\
\resizebox{0.7\textwidth}{!}{
  \includegraphics{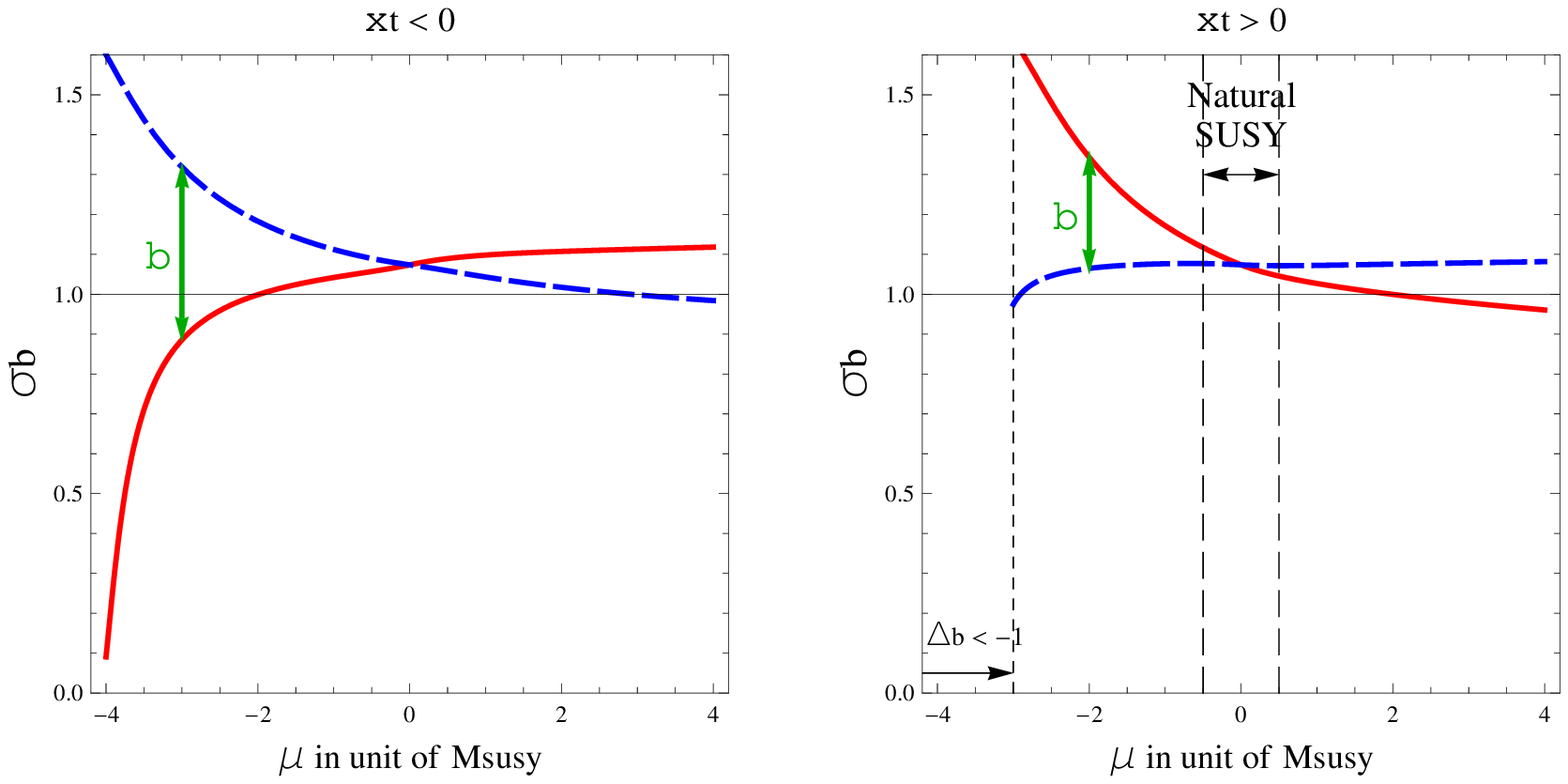}
}\\
\resizebox{0.7\textwidth}{!}{
  \includegraphics{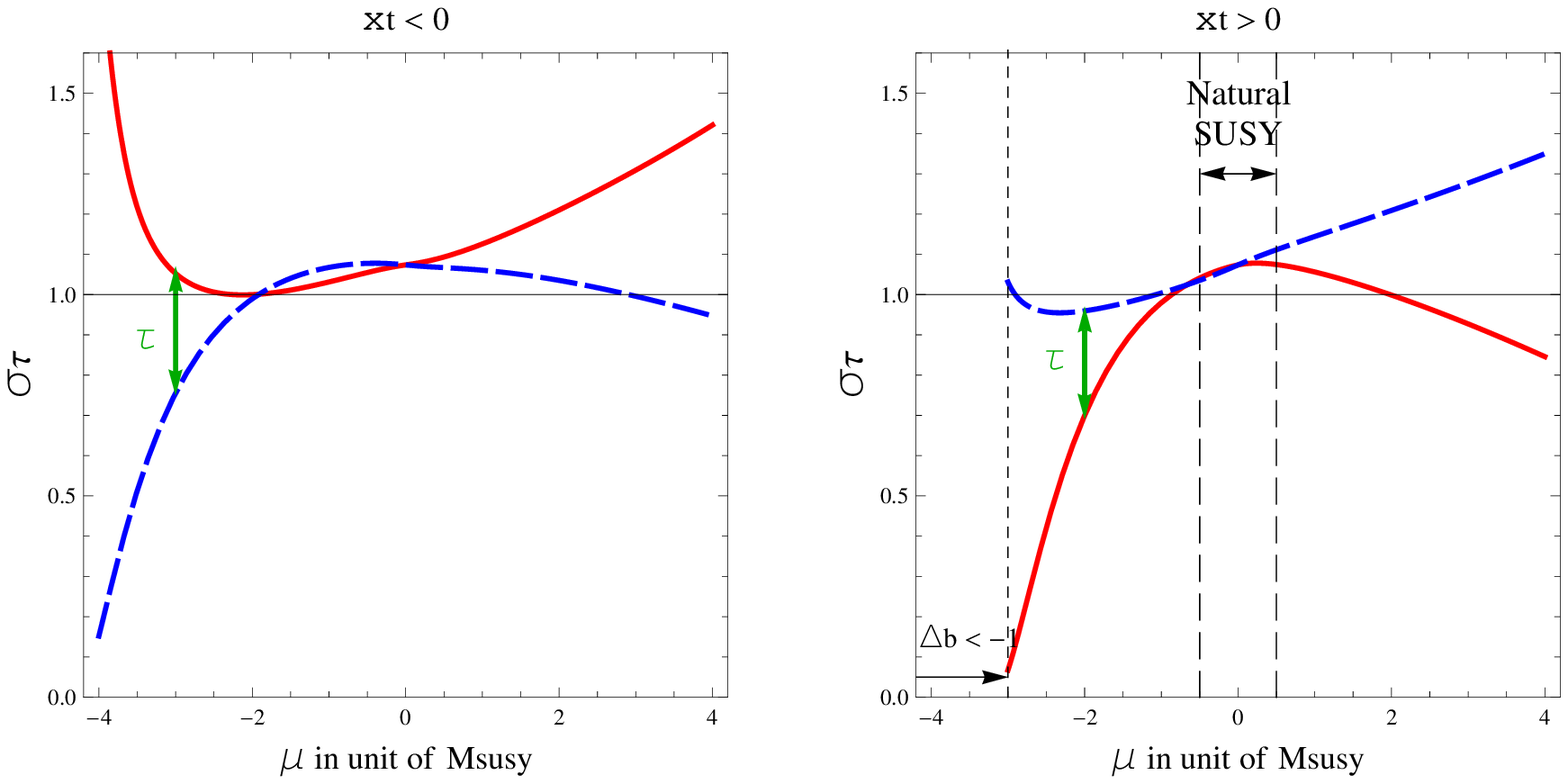}
}
\end{center}
\caption{$\bar\mu$ dependence of
$\sigma_\gamma =\sigma_{\gamma\gamma}/\sigma_{\rm SM}$(upper panel),
$\sigma_b =\sigma_{b\bar b}/\sigma_{\rm SM}$(middle panel), and
$\sigma_b =\sigma_{b\bar b}/\sigma_{\rm SM}$(lower panel) for $m_A=500$~GeV:
Their allowed values are between
the solid red curve (corresponding to $|x_t|=x_{\rm tmax}$) and
the dashed blue curve (corresponding to $|x_t|=x_{\rm tmin}$).
Left(Right) panels show negative(positive) $x_t$ region.
Deviations from unity are enlarged for a large negative $\bar\mu$,
but there the perturbative calculation is unreliable due to a large quantum
correction.}
\label{fig1}
\end{figure}

By using Eq.~(\ref{eq16}) the Higgs mixing angle $\alpha$ is determined from
\begin{eqnarray}
t_{2\alpha} &=& \frac{2M_{12}^2}{M_{22}^2-M_{11}^2}
\simeq \frac{(m_A^2+M_Z^2)s_{2\beta} - 2\Delta M_{12}^2}{
(m_A^2-M_Z^2)c_{2\beta}+(\Delta M_{11}^2-\Delta M_{22}^2)}\ ,
\label{eq21}\\
\Delta M_{12}^2 &\simeq& -\frac{\bar\mu}{s_\beta^2}\ x_t(1-\frac{x_t^2}{6})\ 558{\rm GeV^2} +24\cdot\frac{20}{{\rm tan}\beta}{\rm GeV}^2\ .
\label{eq22}
\end{eqnarray}
Defining $z(\equiv M_Z^2/m_A^2)$, $\delta (\equiv \Delta M_{12}^2/m_A^2)$,
and $\eta (\equiv \frac{1}{2}(\Delta M_{11}^2-\Delta M_{22}^2)/m_A^2)$,
$\epsilon$ is simply given in the first order of $z$, $\delta$, and $\eta$ by
\begin{eqnarray}
\epsilon &=& -2\frac{z+\eta}{{\rm tan}\beta}+\delta\ .
\label{eq22A}
\end{eqnarray}
We note that
$r_{bb}^h$ is related to $\epsilon$ through Eq.~(\ref{eq12}).
With the $x_t$ constraint in Eq.~(\ref{eq20}),
we can derive the allowed region of $r_{bb}^h$ for each $\bar\mu$-value.
Correspondingly, the allowed regions of
$\sigma_\gamma(=\sigma_{\gamma\gamma}/\sigma_{\rm SM}\ =\frac{1}{0.6(r_{bb}^h)^2+0.4})$,\
$\sigma_b(=\sigma_{b\bar b}/\sigma_{\rm SM}\ =\frac{(r_{bb}^h)^2}{0.6(r_{bb}^h)^2+0.4})$
and
$\sigma_\tau(=\sigma_{\tau\tau}/\sigma_{\rm SM}\ =\frac{(r_{\tau\tau}^h)^2}{0.6(r_{bb}^h)^2+0.4})$
are given respectively by the two curves in
Fig.~\ref{fig1} where we take tan$\beta=50$.

The condition $r_{bb}^h=1$, or equivalently $\epsilon = 0$,
$t_{2\alpha}=t_{2\beta}$,
defines the boundary that separates the $\gamma\gamma$ enhancement
and suppression in the parameter space.
\begin{eqnarray}
r_{bb}^h = 1 &\Leftrightarrow &  \epsilon=0 \Leftrightarrow
 \Delta M_{12}^2 = M_Z^2 s_{2\beta}-\frac{\Delta M_{11}^2-\Delta M_{22}^2}{2}t_{2\beta},
\label{eq23}
\end{eqnarray}
This condition is independent of $m_A$ and the quantum correction $\Delta_b$.\\
$\Delta M_{12}^2 > M_Z^2 s_{2\beta} - \frac{\Delta M_{11}^2-\Delta M_{22}^2}{2}t_{2\beta}$ gives $b\bar b$ reduction.
Flavor-tuning (FT) with small $\alpha$ requires a cancellation of $(M_{12}^{2})^{\rm tree}$
by the loop-level $\Delta M_{12}^2$ contribution, which
requires rather large values of $\bar\mu$ and tan$\beta$.
This possibility was raised in ref.\cite{CarenaM12}.

The region of $\gamma\gamma$ enhancement does not overlap with
the region $|\bar\mu|<0.5$ of Natural SUSY for any value of tan$\beta$ from 20 to 60.
For tan$\beta=20$,
$|\bar\mu | \stackrel{>}{\scriptstyle \sim} 2$ is necessary for
$\gamma\gamma$ enhancement.

We give a benchmark point of the FT model in the MSSM (FT1) and
two benchmark points of Natural SUSY (NFT1,NFT2).
\begin{equation}
\begin{array}{cccccccc}
     & \bar\mu & {\rm tan}\beta & x_t & m_h({\rm GeV})
& \sigma_\gamma & \sigma_b & \sigma_\tau  \\
FT1  & -3 & 20 & -2.86 & 124 & 1.17 & 0.89 & 1.05  \\
NFT1 & -0.5 & 20 & 2.70 & 125 & 0.84 & 1.11 & 1.04 \\
NFT2 & -0.15 & 20 & 2.70 & 125 & 0.87 & 1.08 & 1.07
\end{array}
\label{eq25}
\end{equation}
where $M_{\rm susy}=1$~TeV and $m_A=0.5$~TeV.  The relevant sparticle
masses are taken commonly with the values given above Eq.~(\ref{eq5}).
The $m_h$ value is predicted by Eq.~(\ref{eq17}).  We also note that the
predicted BF($B_s\to \mu^+\mu^-$) values of these bench mark points
are consistent with the experimental measurement\cite{LHCb:2012ct}
BF($B_s\to \mu^+\mu^-)=\left(3.2
\stackrel{+ 1.4}{\scriptstyle -1.2}_{\rm stat}
\stackrel{+ 0.5}{\scriptstyle -0.3}_{\rm syst}
\right)\times 10^{-9}   $ within 2$\sigma$.

\noindent \underline{Natural SUSY predictions}\ \ \
Natural SUSY always predicts $b\overline{b}$-enhancement and
$\gamma\gamma$ reduction.\cite{note1}
\begin{eqnarray}
\begin{array}{c|c|c|c}
m_A & \sigma_\gamma  &  \sigma_b & \sigma_\tau \\
\hline
m_A\ge 500~{\rm GeV} & 0.82\sim 0.91 & 1.06\sim 1.12 & 1.04\sim 1.08 \\
m_A\ge 1000~{\rm GeV} & 0.95\sim 0.98 & 1.01\sim 1.03 & 1.01\sim 1.02
\end{array}
\label{eq23}
\end{eqnarray}
Here we have taken $|\mu|\le 500$GeV and the other parameters are fixed with the values
given above Eq.~(\ref{eq5}).

\noindent\underline{\it Concluding remarks}\ \ \

We have explored the $\gamma \gamma $, $b\bar b$ and $\tau \tau$ signals in the MSSM, relative to SM, and also in Natural SUSY. In MSSM an enhancement in the diphoton signal of the 125~GeV Higgs boson relative
to the SM Higgs can be obtained in a flavor-tuned model
with $h=H_u^0$ provided that $|\mu|$ is large($~$TeV) and negative.
A $\gamma \gamma$ enhancement is principally due to the reduction of the $b\bar b$ decay
width compared to $h_{SM}$.
The ratios of $WW^*$ and $ZZ^*$ to their SM values are predicted to be the same as that of $\gamma\gamma$. There
is also a corresponding reduction of the $h$ to $\tau\tau$ signal.
The Tevatron evidence of a Higgs to $b\bar b$ signal in W + Higgs
production \cite{Tevatron} does not favor much $b\bar b$ reduction.
The flavor-tuning of the neutral Higgs mixing angle $\alpha$ requires a
large $\mu\sim$TeV and large tan$\beta$.
For small $|\mu|\stackrel{<}{\scriptstyle \sim} 0.5$~TeV of Natural SUSY,
 $\gamma\gamma$-suppression relative to the SM is predicted.
Thus, precision LHC measurements of
the $\gamma\gamma$, W*W, Z*Z  and $b\bar b$ signals of the 125~GeV Higgs boson
can test MSSM models.

\noindent\underline{\it Acknowledgements}

W.-Y. K. and M. I. thank the hospitality of National Center for Theoretical Sciences
and Academia Sinica while this work was completed.
M.I. is also grateful for hospitality at UW-Madison.
This work was supported in part by the U.S. Department of Energy under grants No. DE-FG02-95ER40896 and
DE-FG02-12ER41811.

\nocite{*}

\bibliography{apssamp}

\begin{thebibliography}{00}
\bibitem{ATLAS} ATLAS Collaboration, Phys.~Lett.~B {\bf 716}, 1 (2012) [arXiv:1207.7214[hep-ex]].
\bibitem{CMS} CMS Collaboration, Phys.~Lett.~B {\bf 716}, 30 (2012) [arXiv:1207.7235[hep-ex]].
\bibitem{NS} M.~Dine, A.~Kagan and S.~Samuel, Phys.~Lett.~B {\bf 243}, 250 (1990).\\
  S.~Kelley, J.~Lopez, D.~Nanopoulos, H.~Pois, and K.~Yuan, Nucl.~Phys.~B{\bf 398}, 3 (1993).\\
V.~Barger, M.~Berger, P.~Ohmann, Phys.~Rev.~D{\bf 49}, 4908 (1994).\\
S.~Dimopoulos and G.~Giudice, Phys.~Lett.~B{\bf 357}, 573 (1995) [arXiv:hep-ph/9507282 [hep-ph]].\\
F.~Gabbiani, E.~Gabrielli, A.~Masiero, and L.~Silvestrini,
 Nucl.~Phys.~B{\bf 477}, 321 (1996) [arXiv:hep-ph/9604387 [hep-ph]].\\
A.~Cohen, D.~Kaplan, and A.~ Nelson, Phys.~Lett.~B{\bf 388}, 588 (1996).\\
K.~Chan, U.~Chattopadhyay and P.~Nath, Phys.~Rev.~D{\bf 58}, 096004 (1998).\\
J.~L.~Feng, K.~T.~Matchev, and T.~Moroi, Phys.~Rev.~D{\bf 61}, 075005 (2000) [arXiv:hep-ph/9909334 [hep-ph]]; Phys.~Rev.~Lett.~{\bf 84}, 2322 (2000) [arXiv:hep-ph/9908309 [hep-ph]].\\
R.~Kitano and Y.~Nomura, Phys.~Lett.~B{\bf 631}, 58 (2005) [arXiv:hep-ph/0509039 [hep-ph]].\\
C~Brust, A.~Katz, S.~Lawrence and R.~Sundrum, JHEP {\bf 1203}, 103 (2012) [arXiv:1110.6670 [hep-ph]].\\
S.~Akula, M.~Liu, P.~Nath, and G.~Peim, Phys.~Lett.~B{\bf 709}, 192 (2012) [arXiv:1111.4589 [hep-ph]].\\
R. Essig, E. Izaguirre, J. Kaplan and J. G. Wacker, JHEP {\bf 1201}, 074 (2012) [arXiv:1110.6443 [hep-ph]].\\
M. Papucci, J. T. Ruderman and A. Weiler, JHEP {\bf 1209}, 035 (2012) [arXiv:1110.6926 [hep-ph]].\\
L.~Hall, D.~Pinner, and J.~Rudermn, JHEP {\bf 1204}, 131(2012)\\
S.~King, M.~Muhlleitner and R.~Nevzorov, Nucl.~Phys.~B{\bf 860}, 207 (2012) [arXiv:1201.2671 [hep-ph]].\\
N.~Arkani-Hamed, talk at WG2 meeting, Oct. 31, 2012, CERN, Geneva.\\
H.~Baer, V.~Barger, P.~Huang, JHEP {\bf 1111}, 031 (2011) [arXiv:1107.5581 [hep-ph]].\\
H.~Baer, V.~Barger, P.~Huang, X.~Tata,in JHEP {\bf 1205}, 109 (2012) [arXiv:1203.5539[hep-ph]].\\
H.~Baer, V.~Barger, P.~Huang, A.~Mustafayev, X.~Tata, arXiv:1207.3343 [hep-ph].\\
D.~M.~Ghilencea, H.~M.~Lee, and M.~Park, JHEP {\bf 1207}, 046 (2012) [arXiv:1203.0569 [hep-ph]].\\
J.~Feng and D.~Sanford, Phys.~Rev.~D~{\bf 86}, 055015 (2012) [arXiv:1205.2372 [hep-ph]].\\
L.~Randall, M.~Reece, arXiv:1206.6540 [hep-ph].
\bibitem{BBM} H.~Baer, V.~Barger, and A.~Mustafayev, Phys.~Rev.~D{\bf 85}, 075010 (2012)
[arXiv:1112.3017v3[hep-ph]].
%
\bibitem{tanb} ATLAS collaboration, ATLAS-CONF-2012-094.
\bibitem{stau}
M.~Carena, S.~Gori, N.~R.~Shah, C.~E.~M.~Wagner, L.-T.~Wang, JHEP {\bf 1207}, 175 (2012) [arXiv:1205.5842 [hep-ph]].
\bibitem{hagiwara} K.~Hagiwara, J.~S.~Lee, and J.~Nakamura, JHEP {\bf 1210}, 002 (2012) [arXiv:1207.0802 [hep-ph]].
\bibitem{falk}  D.~Carmi, A.~Falkowski, E.~Kuflik, T.~Volansky, JHEP {\bf 1207}, 136 (2012)  [arXiv:1202.3144 [hep-ph]].
%
\bibitem{BHG} V.~Barger, H.~E.~Logan, G.~Shaughnessy, Phys.~Rev.~D~{\bf 79}, 115018 (2009) [arXiv:0902.0170[hep-ph]].
\bibitem{Chung} D.~Chung et al., Phys.~Rept.~{\bf 407}, 1-203 (2005). arXiv:hep-ph/0312378
\bibitem{Hunter} John F. Gunion, Howard E. Haber, Gordon Kane, Sally Dawson,
   "The Higgs Hunter's Guide", (Perseus Books, Boulder Colorado 1990).
\bibitem{Dothers} L.~Hall, R.~Rattazzi, and U.~Sarid, Phys.~Rev.~D {\bf 50}, 7048 (1994).\\
 M.~Carena, M.~Olechowski, S.~Pokorski, and C.~E.~M.~Wagner, Nucl.~Phys.~B{\bf 426}, 269.\\
 D.~Pierce, J.~Bagger, K.~Matchev, and R.~Zhang, Nucl.~Phys.~B{\bf 491}, 3 (1997).
\bibitem{Djouadi} A.~Djouadi, Phys. Rep. {\bf 459}, 1 (2008) [hep-ph/0503173v2].
\bibitem{Langacker} P.~Langacker in "the Review of Particle Physics",
J. Beringer et al. (Particle Data Group), Phys.~Rev.~D{\bf 86}, 010001 (2012)
%
\bibitem{CarenaM} M.~Carena et al.,
JHEP {\bf 1207}, 175 (2012); JHEP {\bf 1203}, 014 (2012).
\bibitem{wss} H.~Baer and X.~Tata, {\it Weak Scale Supersymmetry: From
Superfields to Scattering Events},
(Cambridge University Press, Cambridge, England, 2006).
\bibitem{total} V.~Barger, M.~Ishida, and W.-Y.~Keung, 
 Phys.~Rev.~Lett.~{\bf 108}, 261801 (2012) [arXiv:1203.3456[hep-ph]].
\bibitem{book} "Collider Physics", by V.~D.~Barger and R.~J.~N.~Phillips,
(Perseus Book, Boulder, Cololado 1997).
\bibitem{dilaton} V.~Barger, M.~Ishida, and W.-Y.~Keung, Phys.~Rev.~D{\bf 85}, 015024 (2012).
\bibitem{Kingman1} J.~Chang, K.~Cheung, Po-Yan Tseng, and Tzu-Chiang Yuan, JHEP {\bf 1212}, 058 (2012) [arXiv:1206.5853].
%
%
\bibitem{Heine} A.~Denner, S.~Heinemeyer, I.~Puljak, D.~Rebuzzi, and M.~Spira,
Eur.~Phys.~J.~C{\bf 71}, 1753 (2011) [arXiv:1107.5909[hep-ph]].
\bibitem{hdecay} A.~Djouadi, J.~Kalinowski and M.~Spira, Comput.~Phys.~Commun.~{bf108}:56-74, (1998)
\bibitem{HHM} H.~E.~Haber and R.~Hempfling, Phys.~Rev.~D{\bf 48}, 4280 (1993)
[arXiv: hep-ph/9307201]. 	
%
\bibitem{Hmass} V.~Barger, P.~Huang, M.~Ishida, and W.-Y.~Keung, Phys.~Lett.~B{\bf 718}, 1024 (2013) [arXiv:1206.1777[hep-ph]].
%
\bibitem{CarenaM12} M.~Carena, S.~Mrenna, and C.~E.~M.~Wagner, Phys.~Rev.~D~{\bf 62},
055008 (2000) [arXiv:hep-ph/9907422].

\bibitem{LHCb:2012ct}
  R. Aaij {\it et al.}  [LHCb Collaboration],
  arXiv:1211.2674.
\bibitem{Blumn} K.~Blum, R.~T.~D'Agnolo, Phys.~Lett.~B{\bf 714}, 66 (2012) [arXiv:1202.2364[hep-ph]].
\bibitem{Tevatron} The TEVNPH Working Group for the CDF and D0 Collaboration,
 FERMILAB-CONF-12-065-E . arXiv: 1207.0449[hep-ex].
%
\bibitem{note1}
Here we note that as $\bar\mu =0$, then
$ r_{bb}^h = 1+2M_Z^2/m_A^2 \rightarrow
   \sigma_{\gamma\gamma}/\sigma_{\rm SM} = 1-2.4 M_Z^2/m_A^2$,
independently of tan$\beta$.
The smaller $m_A$ gives the larger suppression of $\sigma_\gamma$\cite{Blumn}.
Then, from the $\gamma\gamma$ deviation from unity,
the $CP$-odd state mass $m_A$ could be estimated.
\end{thebibliography}

\end{document}